\documentclass{article}

\usepackage{multirow}
\usepackage{PRIMEarxiv}
\usepackage{float}
\usepackage{authblk}
\usepackage{amsmath}
\usepackage{enumitem}
\usepackage{cancel}
\usepackage{makecell}
\usepackage{algorithm}
\usepackage{algpseudocode}
\usepackage[dvipsnames]{xcolor}
\usepackage[utf8]{inputenc} 
\usepackage[T1]{fontenc}    
\usepackage{hyperref}       
\usepackage{url}            
\usepackage{booktabs}       
\usepackage{amsfonts}       
\usepackage{nicefrac}       
\usepackage{microtype}      
\usepackage{lipsum}
\usepackage{fancyhdr}       
\usepackage{graphicx}       
\usepackage{subcaption}
\usepackage[titletoc]{appendix}
\usepackage{titlesec}
\usepackage{caption} 
\usepackage[backend=biber, maxbibnames=3, minbibnames=3,style=numeric-comp,sorting=none]{biblatex}
\graphicspath{{media/}}     

\title{Towards joint optimization of stellarator coils and support structures}
\author[1, *]{Lanke Fu}
\author[1]{Alan A. Kaptanoglu}
\affil[1]{Courant Institute School of Mathematics, Computing, and Data Science, New York University, New York, NY}
\affil[*]{Corresponding email: lanke\_fu@outlook.com}

\begin{document}
\maketitle

\begin{abstract}
The support structure is an integral part of the design of nuclear fusion reactors, especially 3D stellarator devices. A practical reactor's coils and support structures must have three competing qualities: an accurate magnetic field for good confinement, sufficient rigidity to protect the brittle high-temperature superconductor (HTS) from damage, and a simple geometry for low-cost construction. In existing devices, the coil geometry is often optimized without knowledge of the support structures' design and the coils' true stress and deformation. The support structures are then placed by hand through repeated finite element analyses (FEA) until engineering requirements are met. This makes the structural design of stellarator coil systems lengthy and labor-intensive. Using new developments in differentiable structural mechanics, we present \texttt{coil-fem}, an open-source software tool that integrates support differentiable FEA into the stellarator coil optimization loop. It enables the integrated optimization of coil geometry and support clamp locations to simultaneously reduce magnetic field errors and stresses in the coil body. We also present the first combined coil-support optimization in the stellarator literature. Using a penalty term based on \texttt{coil-fem}, we produced a coil set with $2.4\times$ lower RMS von Mises stress and similar field error compared to an unoptimized baseline.
\end{abstract}

\keywords{nuclear fusion \and stellarator \and structural mechanics \and optimization \and auto-differentiation \and JAX \and FEM}

\section{Introduction}

Stellarators are attractive three-dimensional (3D) fusion devices that generate a rotational transform with external coils. Stellarators do not require steady-state current drive. Unlike tokamaks, they are not susceptible to the current-driven instabilities that can lead to disruptions~\cite{intro_JET_disruption}. Experimental evidence also shows that stellarators can operate at pressures beyond their linear instability thresholds~\cite{intro_pressure_limit}. Therefore, stellarator power plants are thought to have simpler control and heating systems, lower power recirculation, and higher energy efficiency than tokamaks~\cite{intro_Menard_2011}. However, due to their 3D nature, a stellarator's coil system is substantially more complex than a tokamak's. A tokamak's coil set consists mostly of planar coils, whereas a stellarator's coil set often contains optimized non-planar coils with complex geometry and stress distribution.

A stellarator's coil set must accurately reproduce the plasma magnetic field for good confinement. At the same time, it must protect the brittle HTS conductors from enormous mechanical stress. This includes mega-newtons of Lorentz force per meter and thermal expansion due to over $200\text{K}$ of temperature change. Lastly, the coil and support structure must be simple to construct because both are significant contributors to a stellarator's cost~\cite{support-is-pricy-1, aries-cs-cost, shell-vs-cage-2}. Integrated optimization of the coils and support structure has the potential to find more attractive designs for an economical power plant. 

In all existing stellarator devices, the support structures are placed by hand after coil optimization as a post-processing step. If the mechanical loads are too high, the support structure design is manually refined across multiple rounds of structural FEA~\cite{structure-w7x, structure-cfqs-1, structure-cfqs-2, structure-thea}. This process is lengthy, labor-intensive, and often relies on closed-source commercial tools. Because these tools often do not interface with coil optimization codes, it is challenging to refine coil geometry for low mechanical loads at this stage. In addition, during steady-state operation, the mechanical loads deform the coils from the stress-free state and introduce errors in the magnetic field. Correcting field errors due to these deformations also remains challenging. A tool for integrated coil-support optimization may significantly accelerate iteration rates, improve coil performance, and reduce construction costs. 

Recent work has made significant progress toward developing open-source FEA tools for stellarator coils. Two notable tools are \texttt{Parastell} and \texttt{SBGeom}. Both are parametric geometry and meshing toolkits for stellarator coils and blankets, primarily designed for neutronics studies~\cite{parastell, sbgeom}. Packman et al.~\cite{bayesian} optimized the cross-section shape of a single stellarator coil with a fixed centerline to reduce material use while preserving the von Mises stress. This is a scalar stress measure used to predict the yielding of ductile materials, widely used in the coil literature~\cite{structure-cfqs-2, structure-thea}. Packman et al. uses a Bayesian surrogate trained from commercial FEA tools in the optimization loop. Kaptanoglu 2026~\cite{Kaptanoglu2026-dl} performed the first proof-of-concept study that directly integrates FEA into coil optimization to simultaneously target a coil set's von Mises stress and the magnetic field error. However, both Packman et al. and Kaptanoglu et al. assumed fixed support designs. As we will later show in this paper, the locations of the support structures play an important role in the mechanics of stellarator coils. 

\begin{figure}
    \centering
    \includegraphics[width=0.75\linewidth]{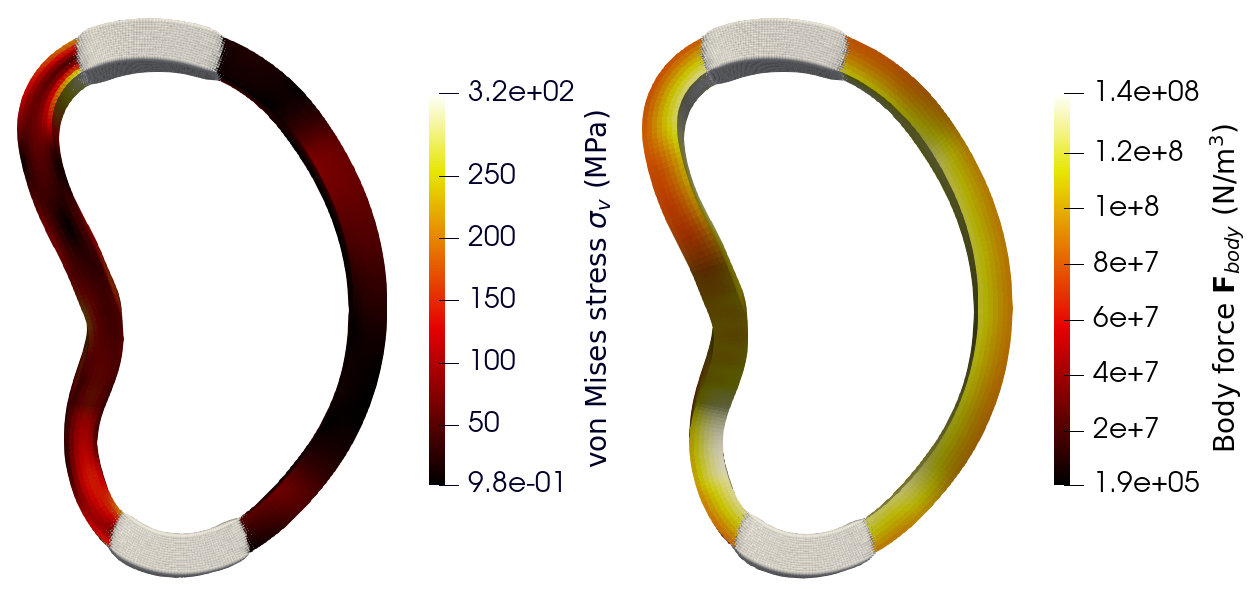}
    \caption{A comparison between the von Mises stress $\sigma_v$ and body force $\mathbf{F}_\text{body}$ on the W7-X type-1 non-planar coil. The gray spheres represent the support structures. Note that while $\mathbf{F}_\text{body}$ is evenly distributed in the coil body, $\sigma_v$ peaks at the left edge of the top support clamp. The data comes from the $n=275$ case in Section \ref{sec:numerical:dolfinx}.}
    \label{fig:vm-vs-f}
\end{figure}

A parallel line of work aims to develop fast, analytic proxies for mechanical stress in coils, such as Lorentz forces~\cite{force-rect}, HTS winding strain~\cite{strain}, torque~\cite{torque}, and stored magnetic energy~\cite{stored-energy}. These proxies have helped reduce HTS damage and support structure costs in recent devices~\cite{stored-energy, coil-epos, coil-csx}. However, none of these proxies model the effect of support structures. As Fig. \ref{fig:vm-vs-f} shows, these metrics cannot fully predict local stress concentrations when support structures are present. Thermal stress, which may be significant in HTS coils that operate at cryogenic temperatures, is also challenging to model without knowledge of the support structures. 

\textbf{Contributions of the present work: }
In this paper, we present \texttt{coil-fem}, the first tool for the combined, gradient-driven optimization of stellarator coils and support structures \cite{coil-fem}. \texttt{coil-fem} is based on \texttt{JAX-FEM}, a recently developed auto-differentiable (AD) finite elements library~\cite{jax-fem-code}. From a set of filamentary coils, it: 1. generates and meshes finite-build coils; 2. generates boundary conditions (BC) based on support clamp locations; 3. computes stress and deformation; and 4. differentiates stress and deformation objectives with respect to coil geometry and support clamp locations. \texttt{coil-fem} is fully differentiable and treats the support structure locations as optimizable variables. Along with \texttt{coil-fem}, we also present the numerical results from the first combined coil-support optimization in the literature. We demonstrate that the placement of the support structure boundary conditions (e.g. clamps in a ``cage'' support structure) is critical for stress reduction. 

This paper is organized as follows. Section \ref{sec:setup} discusses the theory of \texttt{coil-fem}. This includes its finite-build coil model, linear elasticity problem, and support structure model. Section \ref{sec:numerical} benchmarks \texttt{coil-fem} against existing FEA codes and presents numerical results for the combined coil-support optimization on W7-X. Section \ref{sec:conclusions} discusses conclusions and possible future improvements. 

\section{Theory}
\label{sec:setup}
\texttt{coil-fem} solves the following equations for the displacement field $\mathbf{u}$ in each coil's body $\Omega$:
\begin{alignat}{3}
    \nabla \cdot \boldsymbol{\sigma}(\mathbf{u}) + \mathbf{F}_\text{body} &= 0,
        &&\quad \text{in } \Omega
        &&\quad \text{(Linear elasticity)} \label{eq:fem_le} \\
    \boldsymbol{\sigma}(\mathbf{u})\,\mathbf{n} &= -k(\mathbf{x})\,\mathbf{u}.
        &&\quad \text{on } \partial\Omega
        &&\quad \text{(Robin/spring foundation BC)} \label{eq:fem_bc}
\end{alignat}
Eq. \eqref{eq:fem_le} is the linear elasticity equation. Here, $\boldsymbol{\sigma}$ is the Cauchy stress tensor, and $\mathbf{F}_\text{body}$ is the external force density acting on the coils. Eq. \eqref{eq:fem_bc} is a spring foundation boundary condition~\cite{winkler, winkler2}. It approximates the support structure as springs attached to the coil boundary $\partial\Omega$. Here, $k(\mathbf{x})$ is a spatially-varying, isotropic spring coefficient. It is non-zero when the coils are attached to support clamps and zero everywhere else. The spring foundation BC is a penalty approximation of Dirichlet BC commonly used in structural mechanics~\cite{winkler-dirichlet, winkler3-penalty}. It is also the same BC used by Packman et al. and Kaptanoglu et al.~\cite{bayesian, Kaptanoglu2026-dl}. $\mathbf{x}$ and $\mathbf{n}$ are the position and normal vectors on $\partial\Omega$. \texttt{coil-fem} solves and differentiates through Eqs. \eqref{eq:fem_le} and \eqref{eq:fem_bc} using \texttt{JAX-FEM}. Results in this paper use a curved-edge, 10-node tetrahedral (Tet10) mesh and quadratic Lagrange polynomial elements. This is a standard choice for linear elasticity problems in curved geometry~\cite{tet10}.

The following sections will discuss our setup in detail. 
Section \ref{sec:setup:coil} discusses the finite-build coil model and meshing methods.
Section \ref{sec:setup:fem} breaks down the terms and assumptions in Eq. \eqref{eq:fem_le}. 
Section \ref{sec:setup:support} briefly introduces the classification of coil support structures and discusses the construction of Eq. \eqref{eq:fem_bc} and $k(\mathbf{x})$ in more detail.

This paper is a proof of concept with many simplifying assumptions. Tab. \ref{tab:assumptions} provides a list of all assumptions we currently make in \texttt{coil-fem}. For the validity of these assumptions, see Section \ref{sec:setup:coil} - \ref{sec:setup:support}. 

\begin{table}
\centering
\begin{tabular}{|l|l|} 
\hline
Category& Assumptions\\
\hline
\multirow{6}{4em}{Finite-build coils}& - Coil body generated by sweeping a rectangular cross-section along a centerline.\\
& - The cross-sections follow a rotation-minimizing frame.\\
& - Homogeneous and isotropic material properties\\
 & - Uniform current density. (Landreman-Hurwitz self-force.)\\
 &- The mesh is aligned to the coil's cross-sections.\\
 &- Fixed mesh topology during optimization.\\\hline
 \multirow{5}{4em}{Linear elasticity}&- Linear elasticity.\\
 &- Landreman-Hurwitz self-force. \\
 &- Self-force does not depend on deformation.\\
 &- Uniform, isotropic thermal expansion.\\
 &- Small strain decomposition between thermal and other mechanical strains.\\ \hline
 \multirow{3}{4em}{Support structures}&- Cage-type support structure, modeling clamps only (no support beam).\\
 &- Clamps are fixed in space, but their locations are optimizable variables.\\
 &- Spring foundation BC.\\ \hline
\end{tabular}
\caption{List of assumptions for the structural calculations in the present work. Justification and motivation for these assumptions is provided in Sec.~\ref{sec:setup}.}
\label{tab:assumptions}
\end{table}
\subsection{Finite-build coil and meshing}
\label{sec:setup:coil}
Our tool creates the coil body by sweeping a rectangular cross-section along a centerline. The approach is common in multi-filament coil optimization and neutronics calculations~\cite{parastell, sbgeom}. For simplicity, we assume homogeneous material properties, current density, and temperature in this paper. This is only realistic for HTS coils with integrated structural material~\cite{structure-cfqs-1, viper} or copper coils with no support shells. Future versions of \texttt{coil-fem} will support spatially varying material properties, current density, and temperature to accurately model coils with support shells or frames. Another possible addition is to treat material properties, current density, and heating sources/sinks as optimizable variables. This will enable the optimization of conductor and cooling channel placement in the coil body to further reduce loads. 

The orientation of the cross-section during the sweep is given by a moving frame defined around the centerline. Denote the centerline position as $\mathbf{r}_c$. Then, the basis vectors of the moving frame $\{\mathbf{t}, \mathbf{p}, \mathbf{q}\}$ follow a Frenet-Serret type equation:
\begin{equation}\label{eq:frame1}
    \frac{d}{d\phi}
    \begin{pmatrix} \mathbf{t} \\ \mathbf{p} \\ \mathbf{q} \end{pmatrix}
    = \left|\frac{d\mathbf{r}_c}{d\phi}\right|
    \begin{pmatrix} 0 & \kappa_1 & \kappa_2 \\
                    -\kappa_1 & 0 & \kappa_3 \\
                    -\kappa_2 & -\kappa_3 & 0 
    \end{pmatrix}
    \begin{pmatrix} \mathbf{t} \\ \mathbf{p} \\ \mathbf{q} \end{pmatrix},
\end{equation}
Here, $\phi$ is the parametric angle that parameterizes the centerline, and $\{\kappa_1(\phi), \kappa_2(\phi), \kappa_3(\phi)\}$ are periodic functions of $\phi$ that satisfy: 
\begin{equation}\label{eq:frame2}
    \begin{split}
        \mathbf{t} &= \left|\frac{d\mathbf{r}_c}{d\phi}\right|^{-1}\left(\frac{d\mathbf{r}_c}{d\phi}\right),\\
        \kappa\mathbf{n} &=  \kappa_1\mathbf{p}+\kappa_2\mathbf{q} ,  \\
        \mathbf{t}(0) &= \mathbf{t}(2\pi),\\
        \mathbf{p}(0) &= \mathbf{p}(2\pi),\\
        \mathbf{q}(0) &= \mathbf{q}(2\pi).\\
    \end{split}
\end{equation}
Here, $\kappa_1, \kappa_2$ and $\kappa_3$ are the generalized curvatures of the frame. $\kappa$ and $\mathbf{n}$ are the curvature and the normal of the centerline. $\kappa_3$, notably, is the twist of the cross-section:
\begin{equation}
        \kappa_3 =\dfrac{d\alpha}{dl},
\end{equation}
Here, $\alpha$ is the cross-section's alignment angle, and $l$ is the length of the centerline. Typically, one specifies the twist rate $\kappa_3$ by specifying $\alpha$, then evaluates $\kappa_1$ and $\kappa_2$ from the known $\mathbf{r}_c$ using variations of Eqs. \eqref{eq:frame1} and \eqref{eq:frame2}. The choice of $\alpha$ often depends on engineering needs. For example, in the small HTS stellarator EPOS, $\alpha$ is treated as an optimizable variable to reduce winding strain~\cite{coil-epos}. In CFQS, a medium-sized device with copper coils, $\alpha$ is chosen to make the cross-section vertical at the top and bottom to simplify support clamp designs~\cite{structure-cfqs-1}. In this paper, for simplicity, we choose $\alpha$ so that $\{\mathbf{t}, \mathbf{p}, \mathbf{q}\}$ forms a rotation-minimizing frame~\cite{rmf}. 

\texttt{coil-fem} uses a simple, differentiable meshing routine. To mesh a coil with a rectangular cross-section, we parameterize its volume by defining the following coordinates $\{\phi, u, v\}$:
\begin{equation}
            \mathbf{r}(\phi, u, v)
        = \mathbf{r}_c(\phi)
        + \frac{uw_1}{2}\,\mathbf{p}(\phi)
        + \frac{vw_2}{2}\,\mathbf{q}(\phi),
\end{equation}
Here, $w_1$ and $w_2$ are the widths of the cross-section. $\phi\in[0, 2\pi]$ and $u, v\in[-1, 1]$. Then, we use a uniform grid in $\{\phi, u, v\}$ to define the vertices for the mesh. For AD compatibility, \texttt{coil-fem} assumes a fixed mesh topology throughout the optimization. This also allows it to assemble the sparse matrices once and reuse them across optimization steps. While simple, the routine works sufficiently well in practice. Large curvature, torsion, and length lead to high engineering costs and are often penalized during coil optimization. Because of this, a typical stellarator coil set often has mesh-tolerable curvature, torsion, and roughly uniform quadrature spacing. This means even simple, analytic meshing routines can produce sufficiently well-behaved meshes for stellarator coils.


\subsection{Linear elasticity equation}
\label{sec:setup:fem}
In Eq. \eqref{eq:fem_le}, the Cauchy stress tensor is defined as follows: 
\begin{equation}
     \boldsymbol{\sigma} = \lambda \operatorname{tr}(\boldsymbol{\varepsilon})\boldsymbol{I} + 2\mu\, \boldsymbol{\varepsilon},
\end{equation}
where $\lambda$ and $\mu$ are the Lamé parameters, and $\boldsymbol{{\varepsilon}}$ is the strain tensor. A superconducting coil experiences thermal strain due to the $\sim200\text{K}$ temperature change between the stress-free state and the operating state. Our tool treats the thermal strain using small-strain additive decomposition. Assuming isotropic and uniform thermal contraction, the strain tensor is:
\begin{equation}
    \begin{split}
        \boldsymbol{\varepsilon} &= \tfrac{1}{2}\left[\nabla\mathbf{u} + (\nabla\mathbf{u})^\top\right]+(-\Delta l/l) \mathbf{I},
    \end{split}
\end{equation}
where $ (-\Delta l/l)$ is the integrated thermal contraction due to coil cool-down. Future work will implement the full thermo-elastic equations to model the effects of thermal conduction, neutron heating, and cooling channel placement.

The body force density $\mathbf{F}_\text{body}$ in Eq. \eqref{eq:fem_le} consists of the following components:
\begin{equation}
    \mathbf{F}_\text{body} = 
    \rho\mathbf{g}
    + \mathbf{J}\times\mathbf{B}_\text{self} 
    + \mathbf{J}\times\mathbf{B}_\text{mutual}.
\end{equation}
Here, $\mathbf{g}$ is the gravitational acceleration, and $\mathbf{J}$ is the coil current density. We compute the self-field $\mathbf{B}_\text{self}$ using the Landreman-Hurwitz formula for rectangular cross-section coils~\cite{force-rect}. To reduce computation costs, we treat all other coils as 1D filaments when computing $\mathbf{B}_\text{mutual}$. For simplicity, we assume that $\mathbf{F}_\text{body}$ does not depend on the displacement vector $\mathbf{u}$. 

\subsection{Support structure and boundary conditions}
\label{sec:setup:support}

\begin{figure}[htbp]
    \centering
    \begin{subfigure}[t]{0.48\textwidth}
        \centering
        \includegraphics[width=\textwidth, height=6cm, keepaspectratio]{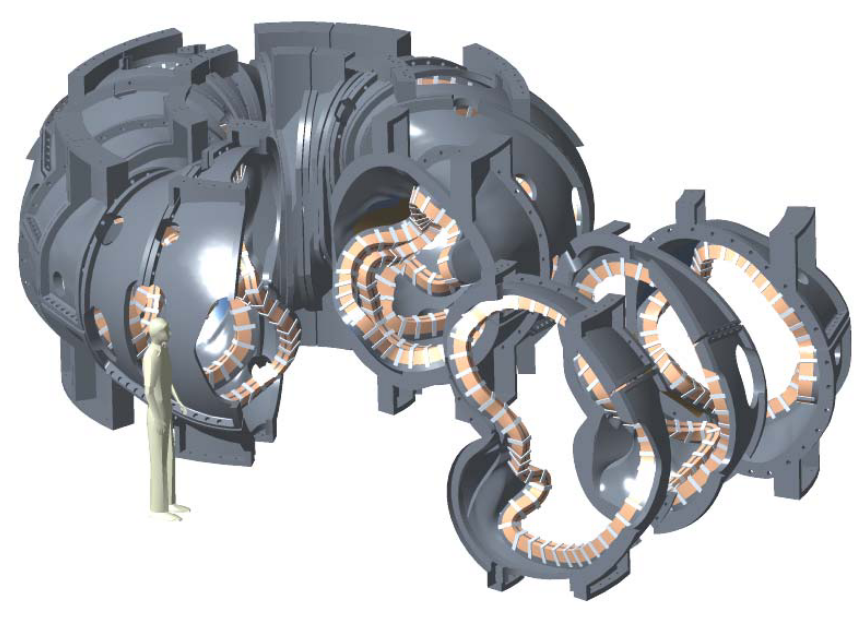}
    \end{subfigure}
    \hfill
    \begin{subfigure}[t]{0.48\textwidth}
        \centering
        \includegraphics[width=\textwidth, height=6cm, keepaspectratio]{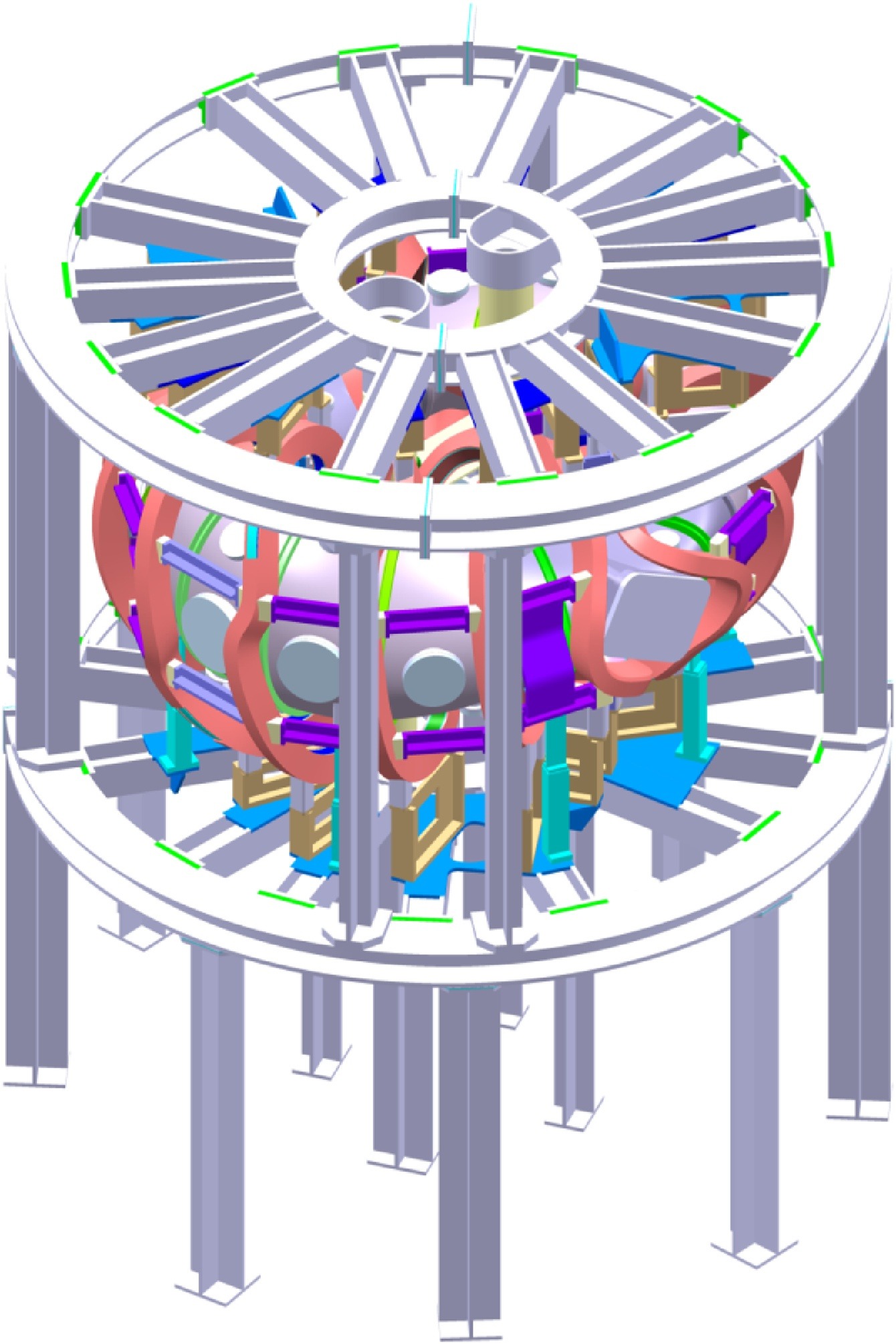}

    \end{subfigure}
    \caption{Left: the shell-type support structure of NCSX~\cite{structure-ncsx-image}. Right: the cage-type support structure of CFQS~\cite{structure-cfqs-2}.}
    \label{fig:support-types}
\end{figure}

There are two main types of stellarator coil support structures: shell-type and cage-type. In a shell-type support structure, the coils are embedded in a rigid shell (Fig. \ref{fig:support-types}, left). The shell may be integrated with the vacuum vessel and blanket for space efficiency. Notable devices with shell-type support include LHD~\cite{structure-LHD} and the canceled NCSX~\cite{structure-ncsx}. In a cage-type support structure, the coils are attached to support frames with a discrete set of clamps (Fig. \ref{fig:support-types}, right). Notable devices with cage-type support include W7-X~\cite{structure-w7x}, HSX~\cite{structure-hsx}, and CFQS~\cite{structure-cfqs-1, structure-cfqs-2}. While no systematic comparison exists, it is believed that a shell-type support can offer greater structural strength at the cost of fabrication complexity. In comparison, a cage-type support offers better plasma access and simpler construction at the cost of high load concentration~\cite{shell-vs-cage-1, shell-vs-cage-2}. An additional advantage of a cage-type support structure is that it may allow a posteriori adjustments to coil positions and orientations to mitigate error fields~\cite{coil-adjustment-1}. 

This paper focuses on the cage-type support structure for its mathematical simplicity and engineering advantages. We model the support structure as a number of fixed support clamps suspended in space. A natural choice of boundary condition for representing this support structure is a Dirichlet BC. However, a true Dirichlet BC is formally singular and will not converge with the mesh resolution~\cite{winkler-dirichlet}. In addition, changing the location of the support clamp will change the node selection in a Dirichlet BC. This will change the structure of the sparse linear system underlying the FEM problem, making the problem AD-incompatible. Therefore, in Eq. \eqref{eq:fem_bc}, we approximate the support clamps with a spring foundation BC instead. When the support clamps move, the spring coefficient $k$ changes, but the structure of the sparse matrix remains the same. This makes Eq. \eqref{eq:fem_bc} AD-compatible and allows \texttt{coil-fem} to reuse sparse matrix structures across iterations.

\begin{figure}
    \centering
    \includegraphics[width=0.5\linewidth]{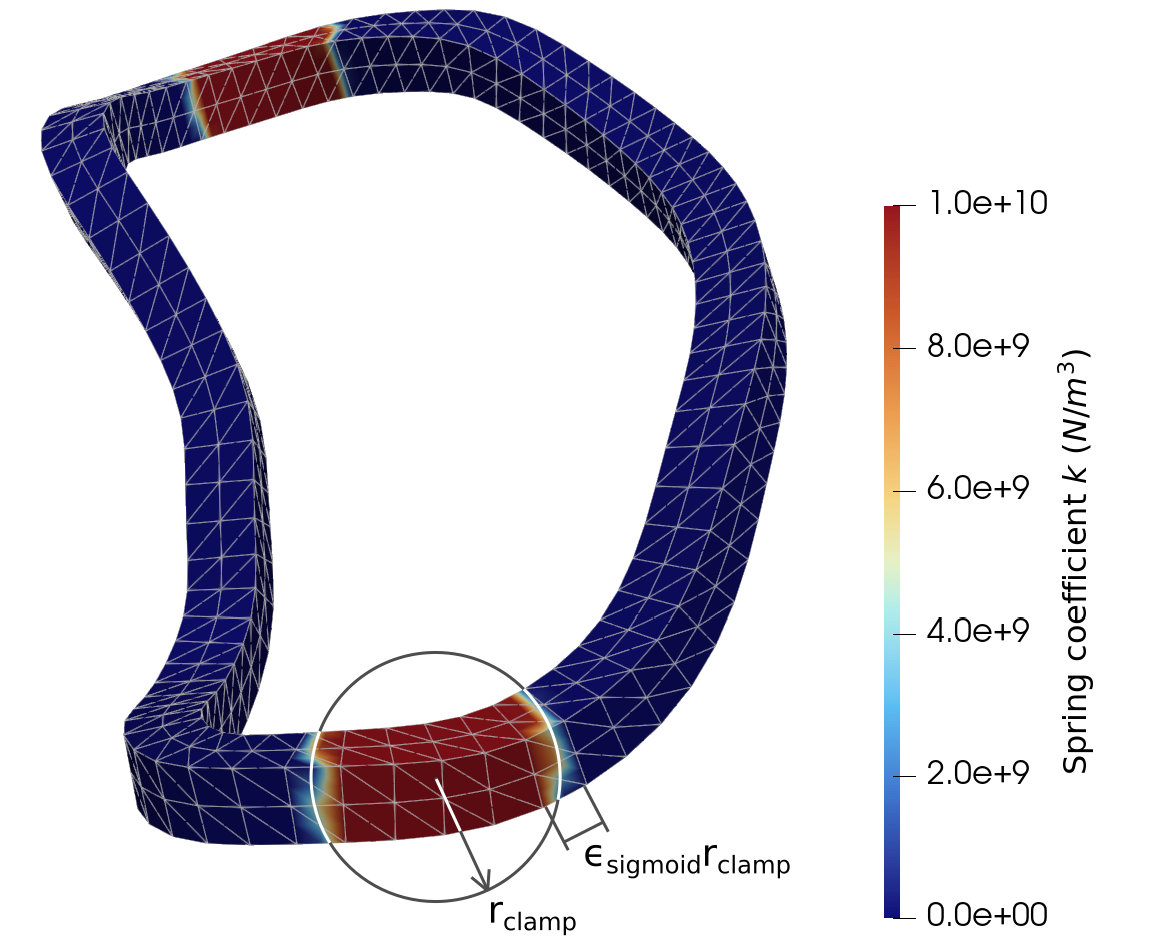}
    \caption{Spring coefficient function $k(x,\phi_i)$ that emulates two support clamps on the top and bottom of a W7-X coil. }
    \label{fig:bc}
\end{figure}

To represent a cage-type support structure, we choose the following $k$:
\begin{equation}\label{eq:k}
    k(\mathbf{x}, \{\phi_i\}) = k_0\sum_{i=1}^{N_\text{cl}} S\left[\frac{r_\text{cl}^2- |\mathbf{x} - \mathbf{r}_{ci}(\phi_i)|^2}{\epsilon_\text{cl}^2r_\text{cl}^2}\right].
\end{equation}
Here, $N_\text{cl}$ is the number of support clamps on each coil. $\{\phi_i\}$ stores the support clamp locations in the coil parametric angle $\phi$. $k_0$ is a constant spring coefficient. $S(x)=1/(1+e^{-x})$ is the logistic sigmoid function. $r_\text{cl}$ is a constant clamp size. $\epsilon_\text{cl}$ is a small, dimensionless width parameter. Fig. \ref{fig:bc} shows a graphical interpretation of Eq. \eqref{eq:k}. Eq. \eqref{eq:k} sets $k$ to $k_0$ at all surface nodes inside $N_\text{cl}$ spheres located at $\mathbf{r}_{ci}(\phi_i)$. Each sphere represents a support clamp with a half-width of approximately $r_\text{cl}$. At the boundary of each sphere, $k$ smoothly transitions from $k_0$ to zero in a sigmoid form. The transition region has a width of $\epsilon_\text{cl}r_\text{cl}$. Outside the spheres, $k=0$. $k(\mathbf{x},\{\phi_i\})$ is analytic for all $\mathbf{x}$ and $\phi_i$. This construction allows us to differentiate through the FEA problem and optimize $\phi_i$ to reduce mechanical loads.

In this paper, the spring coefficient $k$ is isotropic and set to a large, fixed value. It is worth noting that Eq. \eqref{eq:fem_bc} serves as an auto-differentiable alternative to the Dirichlet BC and cannot accurately predict flexing in the support cage. To accurately model support cage flexing, future versions must couple the coil models to a finite-element/beam network model of the support cage or replace $k$ with an anisotropic value obtained by performing sensitivity analyses on the support cage.

\section{Numerical results} \label{sec:numerical}

This section presents two numerical studies that validate \texttt{coil-fem's} accuracy and effectiveness in optimization. Section \ref{sec:numerical:dolfinx} benchmarks \texttt{coil-fem} against \texttt{DOLFINx}, a popular Python FEA suite. Section \ref{sec:numerical:opt} presents a combined coil-support optimization enabled by \texttt{coil-fem}. Our studies are based on a simplified version of the W7-X coil set~\cite{overview-w7x}. A true W7-X coil consists of an Nb-Ti winding pack surrounded by a 316LN stainless steel support case. Here, as a simple proof-of-concept, we use uniform 316LN stainless steel for the entire coil body. Its material properties are listed in Tab. \ref{tab:material}~\cite{material}. The spring coefficient in the boundary condition, $k_0$, is set to $10^{10} \text{ N/m}^3$. The true W7-X support structure consists of 3 types of support elements that link each coil to the Central Support Structure and adjacent coils~\cite{structure-w7x}. Instead, here we define two support clamps for each coil with $r_\text{cl}=0.3\text{ m}$. In the first study, the clamps are located at the top and bottom of each coil. In the second study, the clamps are allowed to move to reduce von Mises stress.
\begin{table}
    \centering
    \begin{tabular}{|c|c|}\hline
 Coil width $w_1, w_2$&$0.2 \text{ m}$\\\hline
         Young's Modulus $E$& $2.05\times10^{11}\text{ Pa}$\\\hline
         Poisson's ratio $\nu$& $0.3$\\\hline
         Density $\rho$ (for gravity)& $8000\text{ kg/m}^{3}$\\\hline
 Integral thermal contraction $(-\Delta l/l)$&$0.29\%$\\\hline
    \end{tabular}
    \caption{Material properties of 316LN stainless steel in conditions relevant for a Nb-Ti coil. }
    \label{tab:material}
\end{table}
\subsection{Benchmark}\label{sec:numerical:dolfinx}

To benchmark \texttt{coil-fem} against \texttt{DOLFINx}~\cite{dolfinx}, we calculate the von Mises stress and deformation of the W7-X type 1 non-planar coil due to $\mathbf{B}_\text{self}$, thermal expansion, and gravity. To observe convergence, we scan the centerline resolution $n$. The total cell-number scales approximately by $O(n^3)$. For each value of $n$, we perform two \texttt{DOLFINx} runs with different implementations of $\mathbf{B}_\text{self}$. Case 1 uses the same Landreman-Hurwitz formula as \texttt{coil-fem}. This serves to test our \texttt{JAX-FEM} linear elasticity setup. Case 2 finds $\mathbf{B}_\text{self}$ by directly integrating current elements over the coil volume. This serves to test our implementation of the Landreman-Hurwitz formula. For consistency, the \texttt{DOLFINx} runs use the same Tet10 mesh and second-order Lagrange basis as \texttt{coil-fem}. 

\begin{figure}
    \centering
    \includegraphics[width=0.85\linewidth]{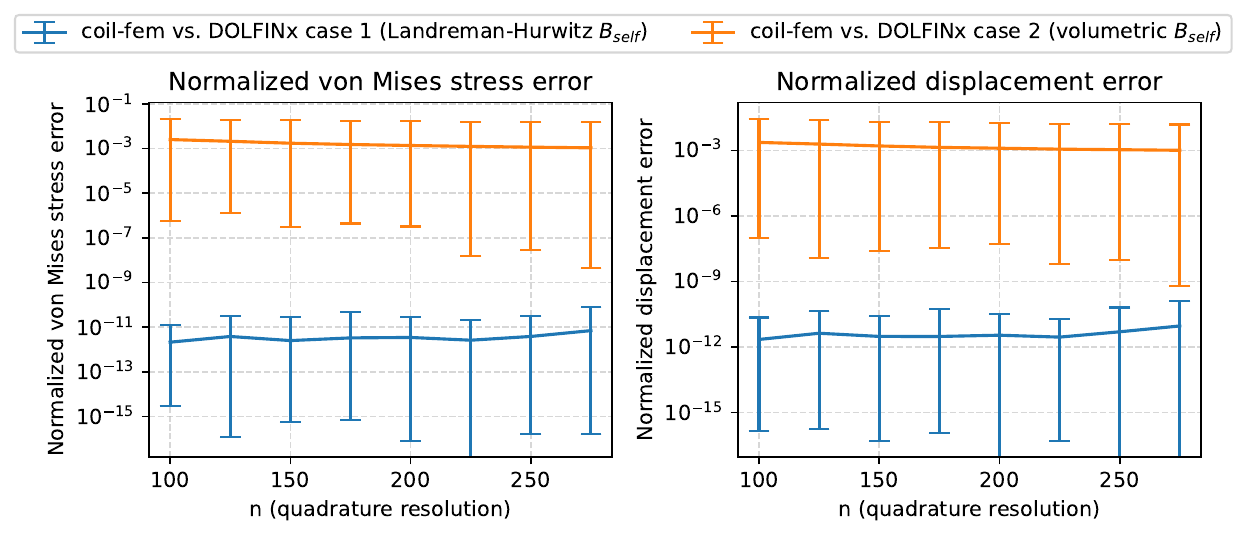}
    \caption{The minimum, maximum and median point-wise errors in von Mises stress (left) and displacement (right) between \text{coil-fem} and \texttt{DOLFINx}. The values are normalized by the \text{coil-fem} means.}
    \label{fig:bench:error}
\end{figure}
\begin{figure}
    \centering
    \includegraphics[width=0.85\linewidth]{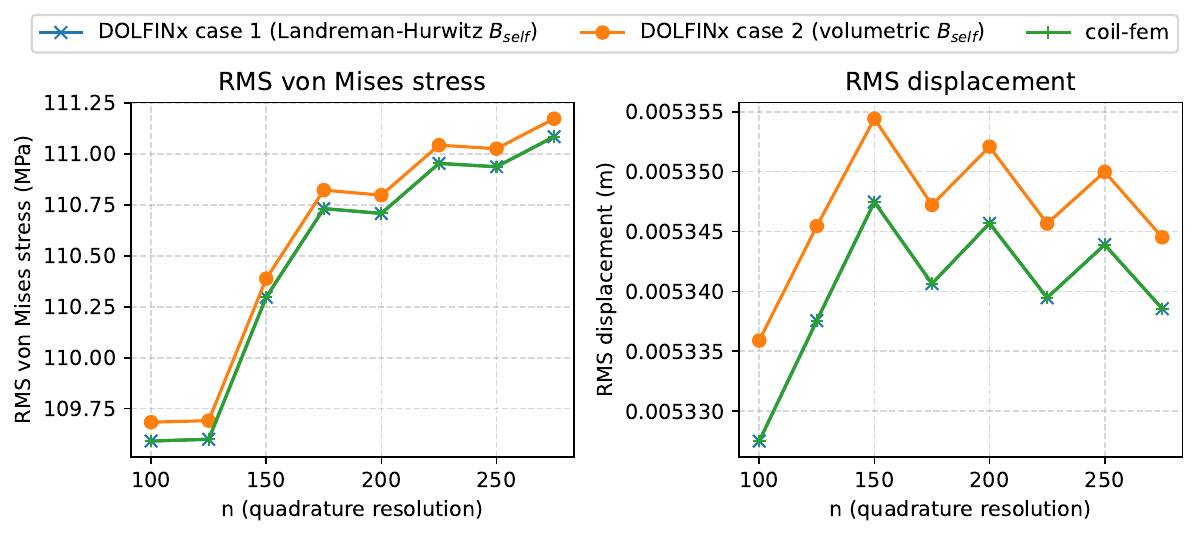}
    \caption{The RMS von Mises stress and displacements from \texttt{coil-fem} and \texttt{DOLFINx}. Note the apparent convergence of all three cases.}
    \label{fig:bench:conv}
\end{figure}

Fig. \ref{fig:bench:error} shows the point-wise error between \texttt{coil-fem} and \texttt{DOLFINx}, normalized by the \texttt{coil-fem} means. \texttt{coil-fem} agrees with \texttt{DOLFINx} case 1 within $10^{-8}$. This suggests that our linear elasticity problem is correctly implemented. \texttt{coil-fem} agrees with \texttt{DOLFINx} case 2 within $5\%$, with the error decreasing as $n$ increases. This shows that our implementation of the Landreman-Hurwitz $\mathbf{B}_\text{self}$ correctly approaches the volume integral at high resolutions. Fig. \ref{fig:bench:conv} compares the RMS von Mises stress and displacements among all three cases. In all cases, the RMS values agree within $2\%$ and show some level of convergence. 

\begin{figure}
    \centering
    \includegraphics[width=0.6\linewidth]{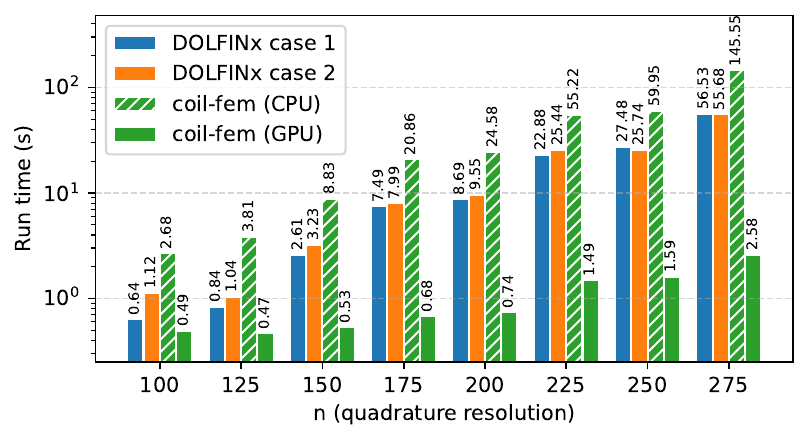}
    \caption{The forward-solve time of \texttt{coil-fem} and \texttt{DOLFINx}. The \texttt{DOLFINx} runs use 4 Intel Xeon Platinum 8592+ 64C CPUs. The \texttt{coil-fem} runs use one Nvidia L40S GPU.}
    \label{fig:bench:time}
\end{figure}

\texttt{coil-fem} supports both CPU and GPU. The CPU backend uses \texttt{PETSc}, the same sparse solver that \texttt{DOLFINx} uses by default~\cite{petsc}. The GPU backend uses \texttt{cuDSS} and \texttt{spineax}~\cite{cudss, spineax}, and performs the full computation, from meshing to differentiation, entirely on GPUs. Fig. \ref{fig:bench:time} benchmarks \texttt{coil-fem} on CPU and GPU. The figure also includes the runtime of \texttt{DOLFINx} on CPU for reference. Note that the \texttt{coil-fem} runtime includes meshing, $\mathbf{B}_\text{self}$, and FEA, while the \texttt{DOLFINx} runtime includes FEA only. As the figure shows, GPU \texttt{coil-fem} outperforms the rest by orders of magnitude. CPU \texttt{coil-fem} is the slowest among all cases. Its runtime is comparable to CPU \texttt{DOLFINx} at low $n$ but increases faster as $n$ increases. This is possibly due to the additional assembly that \texttt{coil-fem} performs for adjoint differentiation.

\subsection{Optimization}\label{sec:numerical:opt}

This section demonstrates the effectiveness of \texttt{coil-fem} in a coil optimization loop. For completeness, we compare the results from 4 coil problems, labeled A to D:
\begin{enumerate}[label=\Alph*.]
\item Minimizing a \texttt{coil-fem} load penalty, varying the clamp locations only.
\item Minimizing a \texttt{coil-fem} load penalty, varying the coil $x, y, z$ Fourier coefficients only.
\item Minimizing a \texttt{coil-fem} load penalty, varying both the coil Fourier coefficients and the clamp locations.
\item A control case that minimizes a coil-force penalty by varying the coil Fourier coefficients.
\end{enumerate}
The coil optimization uses the penalty:
\begin{equation}
\label{eq:filament_stage_2}
    J \equiv 10^{12}\max(J_B-J_{B,\text{baseline}}, 0)^2+ J_\text{load}  + 200 J_{l} + 100 J_{\text{cc}} + 100J_{\text{cs}} + 10J_\text{link} + 1000J_\kappa.
\end{equation}
This function contains 7 penalty terms: the field error $J_B$, the mechanical load term $J_\text{load}$, the total length term $J_l$, the coil-coil distance $J_\text{cc}$, the coil-plasma distance $J_\text{cs}$, the linking term $J_\text{link}$, and the curvature term $J_\kappa$. The penalty weights are empirically chosen based on the typical values of $J_B$ through $J_\kappa$. Except for $J_\text{load}$, all other penalty terms are standard in the stellarator coil literature~\cite{filament1, filament2, filament3}. Their definitions are listed in Appendix \ref{app:terms}. For problems A, B, and C, $J_\text{load}$ is the volume integral of the squared von Mises stress $\sigma_v$ calculated with \texttt{coil-fem}:
\begin{equation}
    J_\text{load} =10^{-18} \int dV  \sigma_v^2,
\end{equation}
For problem D, $J_\text{load}$ is the L2 Lorentz force objective~\cite{force-rect}:
    \begin{equation}
        J_\text{load} = (2.5\times10^{-5})\frac{1}{2}\sum_i\frac{1}{l'_i}\left[\int_{C_i} \text{max}(|F'| - F_0 , 0)^2 dl'_i\right],
    \end{equation}
where $\int_{C_i}$ integrates over the centerline $C_i$, $F'$ is the centerline Lorentz force density, and $l'$ is the length of the coil. The target force $F_0$ is set to zero to minimize coil forces as much as possible. We use the standard "EIM" vacuum configuration as the target magnetic field~\cite{w7x-standard}. All optimizations are cold-started from circular initial conditions. As a baseline case, we compare all optimization results with the original W7-X coils, supported by clamps located at the top and bottom. As discussed at the start of Section \ref{sec:numerical}, the number and locations of support clamps in this study are distinct from those in the actual W7-X device. Our choice to conduct this study with only two clamps per coil is purely for simplicity and ease of comparison.

To evaluate $\sigma_v$, we run GPU \texttt{coil-fem} with a resolution of $n=80$, which corresponds to $1920$ mesh cells per coil. We perform the optimization with the L-BFGS algorithm in 3 multi-grid steps \cite{multigrid} using \texttt{Simsopt} and \texttt{SciPy}~\cite{simsopt, 2020SciPy-NMeth}. Tab. \ref{tab:opt:time} shows the runtime statistics of all 4 problems. As the table shows, each $\nabla J$ evaluation requires about $10\text{s}$. For reference, each $\nabla J$ evaluation requires about $6\text{s}$ in \texttt{StellCoilBench}, which performs finite differencing over multiple \texttt{DOLFINx} runs with similar cell counts, first-order Lagrange basis, and one quadrature point per cell~\cite{Kaptanoglu2026-dl}. 

\begin{table}
    \centering
    \begin{tabular}{|l|c|c|c|c|c|}\hline
         &   Num. DOFs &Tot. wall time&  \makecell{Tot. L-BFGS\\iterations}&  \makecell{Tot. $\nabla J$\\evaluations}&\makecell{Time per $\nabla J$\\evaluation}\\\hline
         \makecell[l]{A: Fixed (W7-X) coils, \\optimized supports}& $10$           &$0.28 \text{ h}$&  $58$  &  $102$ &$10.20\text{ s}$\\\hline
         \makecell[l]{B: Optimized coils, \\fixed supports}&        $136, 196, 256$&$3.04 \text{ h}$&  $313$ &  $1073$& $9.41\text{ s}$\\\hline
         \makecell[l]{C: Optimized coils, \\optimized supports}&    $146, 206, 266$&$6.48 \text{ h}$&  $1500$&  $2472$& $9.42\text{ s}$\\\hline
         \makecell[l]{D: Low-force coils, \\fixed supports}&        $136, 196, 256$&$0.27 \text{ h}$&  $1500$&  $3022$& $0.32\text{ s}$\\\hline
    \end{tabular}
    \caption{Runtime statistics of cases A through D. The optimization is run on an L40S GPU in 3 Fourier multi-grid steps. Each step permits $500$ maximum L-BFGS iterations.}
    \label{tab:opt:time}
\end{table}

\begin{figure}
    \centering
    \includegraphics[width=1\linewidth]{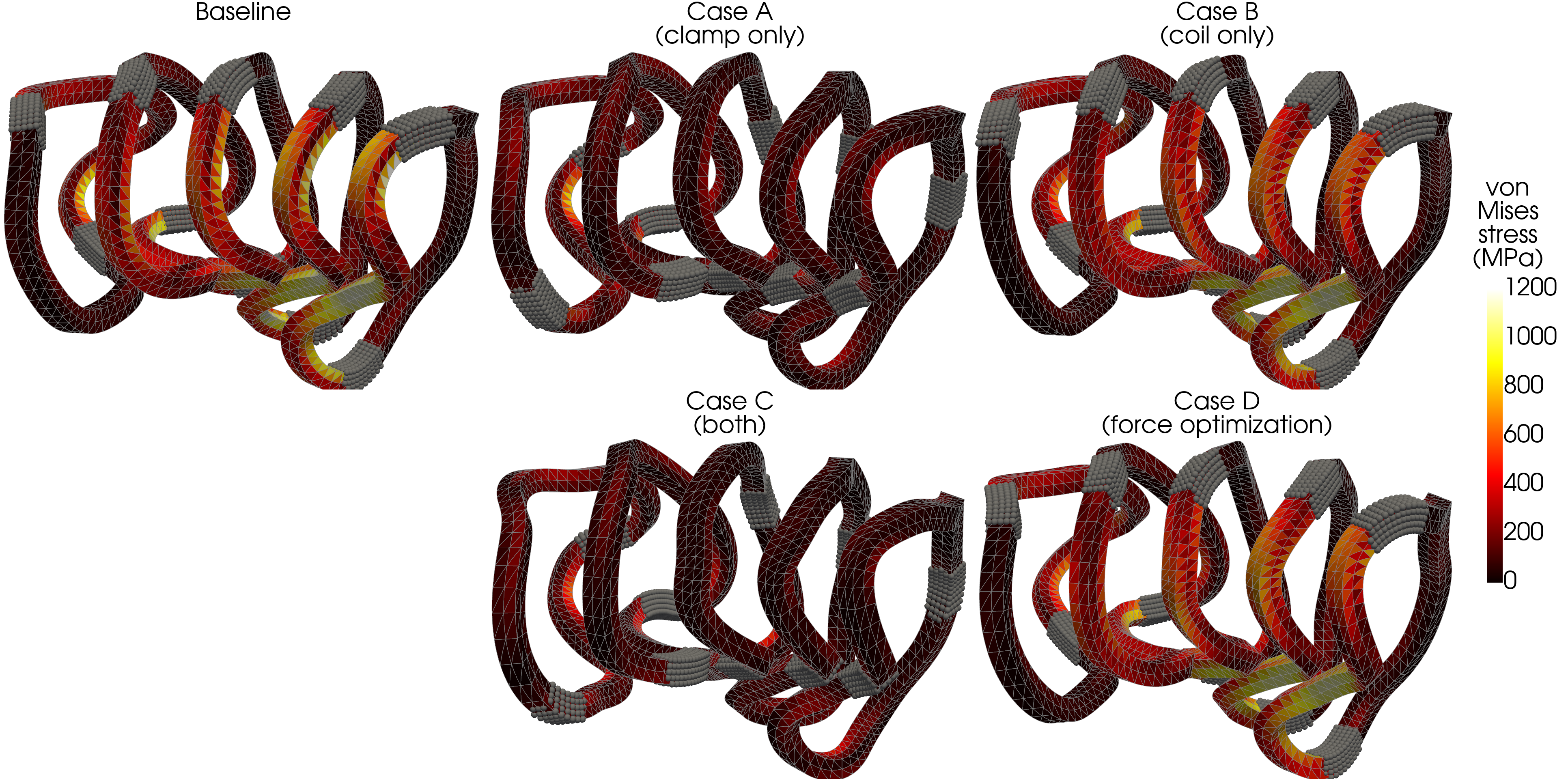}
    \caption{A comparison of $\sigma_v$ (color) and support clamp locations (gray glyphs) across all 5 configurations. Here, case B and D have the same clamp locations as the baseline. In case A and C, the support clamp locations are optimized. This leads to substantial reductions in peak and RMS $\sigma_v$.}
    \label{fig:3d}
\end{figure}

\begin{figure}
    \centering
    \includegraphics[width=0.95\linewidth]{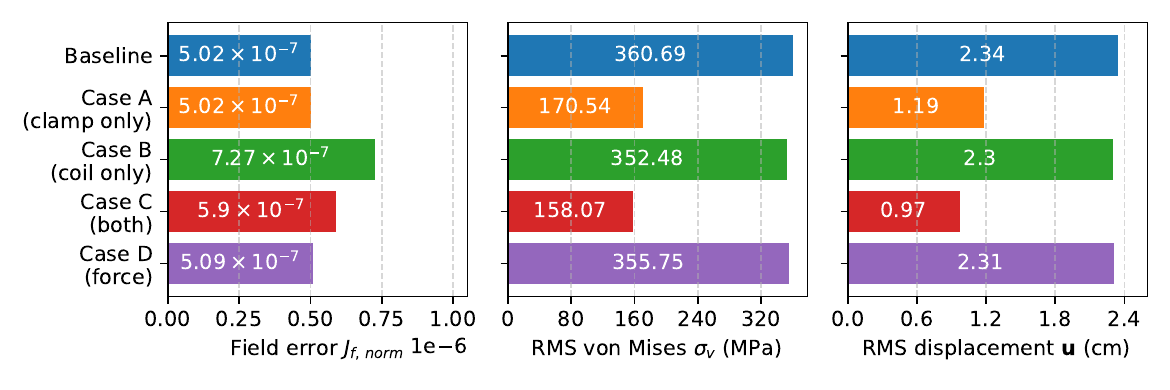}
    \caption{Global metrics of all 5 configurations. Note that cases A and C have significantly reduced $\sigma_v$ and $\mathbf{u}$. This highlights the importance of clamp locations when modeling coil stress and deformation.}
    \label{fig:opt:overview}
\end{figure}

We now summarize our findings from the optimization results. Fig. \ref{fig:3d} compares the support locations and $\sigma_v$ distributions in all 5 cases. Fig. \ref{fig:opt:overview} compares the field error, $\mathbf{F}_\text{body}$, and $\sigma_v$ in all 5 cases. Compared to the baseline, cases A and C reduce $\sigma_v$ by over $2\times$ by optimizing the locations of the support clamps. In comparison, as case B shows, optimizing the coil geometry with fixed support clamp locations only leads to marginal improvements in $\sigma_v$. This confirms the important role of support clamp locations in the structural mechanics of stellarator coils. Case D shows that optimizing the coil forces does not always lead to reduced $\sigma_v$, justifying the integration of structural FEA into stellarator coil optimization. Overall, case C outperforms the rest, offering nearly $2.4\times$ reductions in both RMS $\sigma_v$ and $\mathbf{u}$ over the baseline. This demonstrates the effectiveness of \texttt{coil-fem} as a tool for combined coil-support optimization. 

\begin{figure}
    \centering
    \includegraphics[width=\linewidth]{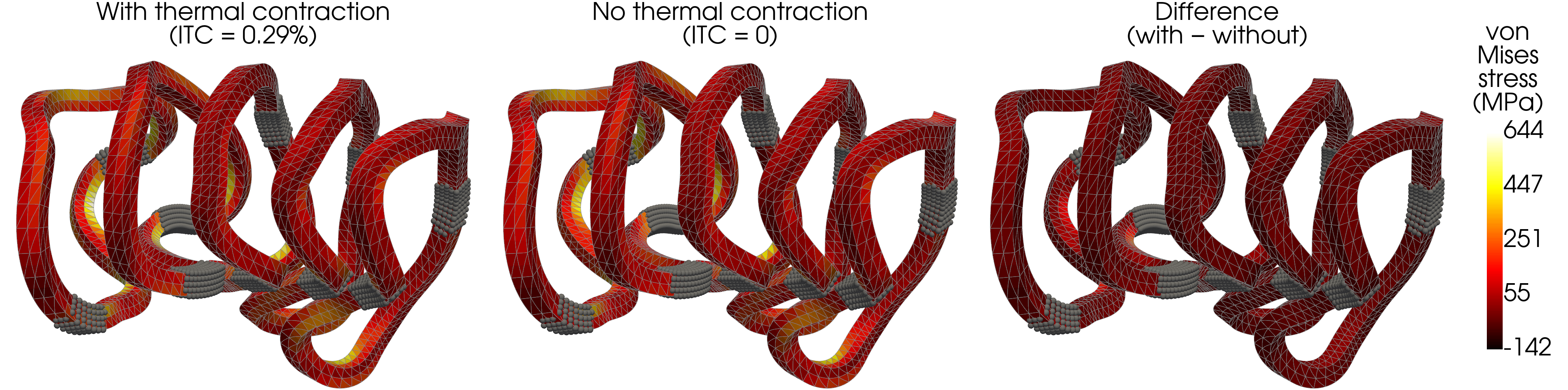}
    \caption{The values of $\sigma_v$ in case C (optimizing both the clamp locations and coils) with (left) and without thermal contraction (middle). The right figure shows their differences.}
    \label{fig:itc-or-not}
\end{figure}

The dominant contributor to the stress $\sigma_v$ in a W7-X-scaled coil set is likely the Lorentz force. The contributions of gravitational force are negligible. The gravitational force density acting on a 316LN stainless steel coil body is $\sim80\text{kN}/\text{m}^3$. In comparison, the typical Lorentz force density is $\sim 100\text{MN}/\text{m}^3$, three orders of magnitude higher. The contributions from thermal contraction $(-\Delta l/l)$ are secondary but still significant. Fig. \ref{fig:itc-or-not} shows the changes in $\sigma_v$ distribution in case C when thermal contraction is neglected. The RMS $\sigma_v$, which our optimization targets, shifts by $3.6\%$ from $152.9\text{MPa}$ with thermal contraction to $147.5\text{MPa}$ without thermal contraction. The maximum $\sigma_v$, which constrains feasibility, shifts by $17.0\%$ from $643.4\text{MPa}$ with thermal contraction to $534.2\text{MPa}$ without thermal contraction. While not an order-of-magnitude difference, it can still impact the feasibility of a coil design. The relative importance of Lorentz force and thermal contraction will likely shift for devices with different coil sizes and field strengths. The presence of heat sources, sinks, and temperature gradients may also introduce additional stress concentrations not present in a model with uniform thermal contraction. The significance of these effects requires further investigation.

\section{Conclusions and future work}
\label{sec:conclusions}

This paper presents \texttt{coil-fem}, a fully differentiable FEA tool for optimizing stellarator coils and supports \cite{coil-fem}. We also present the first combined coil-support optimization in the literature. Overall, \texttt{coil-fem} is an effective tool for reducing mechanical loads in stellarator coils. Although slower than analytic proxies, it can model the effect of changing support locations, which is impossible with existing proxies. 

In future work, we plan to extend \texttt{coil-fem} to support coil models with higher fidelity. Possible improvements include spatially varying material properties, current, and a full thermo-elastic system with realistic thermal expansion curves, heat sources, sinks, and temperature gradients. These improvements can enable the optimization of conductor and cooling system placements in the coil body to minimize costs and maximize rigidity. Future versions will also introduce realistic support structure models. A finite element model is likely needed for shell-type supports, and a beam network model is likely needed for cage-type supports. In addition to load reduction, we believe \texttt{coil-fem} can also be used for error field mitigation. Using the displaced coil geometry produced by \texttt{coil-fem}, one can minimize a coil set's error field due to its deformation under mechanical loads. This can lead to coil designs with improved field accuracy and particle confinement. Lastly, applying \texttt{coil-fem} to stellarator coil databases may also lead to new coil mechanics proxies with improved speed and accuracy. These can take the form of learned surrogates~\cite{surrogate1, surrogate2, surrogate3} or analytic models like Euler, Kirchhoff, or Cosserat-type rods commonly used in structural mechanics.

\section*{Acknowledgments}

This work is supported by the Simons Hidden Symmetry Collaboration under award 560651.

\section*{Data availability statement}

The data that supports the findings of this study is openly available \cite{zenodo-data}. 

\appendix
\numberwithin{equation}{section}
\renewcommand{\theequation}{\thesection\arabic{equation}}
\section{Filament penalty terms}
\label{app:terms}
The definitions of all terms (except for $J_\text{load}$) in Eq. \eqref{eq:filament_stage_2} are:
\begin{enumerate}
    \item $J_B$, the normalized squared flux:
\begin{equation}
    \label{eq:JB}
    J_B \equiv \frac{1}{2} \frac{\int_S\left(B_\text{norm} -B_T\right)^2 dS}{\int_S|\mathbf{B}|^2 dS}.
\end{equation}
Here, $B_\text{norm}$ is the coil normal field at the plasma boundary. $B_T$ is the target normal field and is zero in a vacuum configuration. $\int_{S}dS$ is the plasma surface integral. The target $J_{B,\text{baseline}}$ in Eq. \eqref{eq:filament_stage_2} is the $J_B$ of the baseline W7-X coils.
    \item $J_l$, the maximum coil length constraint:
    \begin{equation}
        J_l = \frac{1}{a^2}\sum_{i = 1}^{N_\text{coil}}0.5\max(l_i - l_{i, \text{baseline}}, 0)^2.
    \end{equation}
    Here, $l_{i, \text{baseline}}$ is the length of the $i$-th W7-X coil. $a$ is the plasma minor radius.
    \item  $J_{\text{cc}}$, the coil-coil spacing constraint:
    \begin{equation}
            J_\text{cc} \equiv \frac{1}{d^{\min}_{\text{cc}}}\sum_{i = 1}^{N_\text{coil}} \sum_{j = 1}^{i-1} \int_{C_i} \int_{C_j} \max(d^{\min}_{\text{cc}} - \| \mathbf{r}_{c,i} - \mathbf{r}_{c,j} \|_2, 0)^2 ~dl_j ~dl_i,\\
    \end{equation}
Here, $\mathbf{r}_{c,i}, \mathbf{r}_{c,j}$ are locations on pairs of coil centerlines, $C_i$ and $C_j$. $\int_{C_i}dl_i$ and $\int_{C_j}dl_j$ represent integrals along coil centerlines. The threshold for the minimum coil-coil distance, $d^{\min}_{\text{cc}}$, is set based on W7-X measurements.
    \item $J_{\text{cs}}$, the coil-plasma spacing constraint:
    \begin{equation}
        J_{\text{cs}} \equiv \frac{1}{a} \sum_{i = 1}^{N_\text{coil}} \int_{C_i} \int_{S} \max(d^{\min}_{\text{cs}} - \| \mathbf{r}_{c,i}  - \mathbf{s} \|_2, 0)^2 ~dl_i ~dS.
    \end{equation}
Here, $\mathbf{s}$ represents locations on the plasma surface. The threshold for the minimum coil-plasma distance, $d^{\min}_{\text{cs}}$, is set based on W7-X measurements.
    \item $J_{\text{link}}$, the linking number constraint. Each pair of coils $C_1$ and $C_2$ contributes:
    \begin{equation}
        \text{Link}(C_1, C_2) = \frac{1}{4\pi} \left| \oint_{c_1}\oint_{c_2}\frac{\mathbf{r}_{c, 1} - \mathbf{r}_{c,2}}{|\mathbf{r}_{c, 1} - \mathbf{r}_{c,2}|^3} \cdot(d\mathbf{r}_{c,1}  \times d\mathbf{r}_{c,2} ) \right|.
    \end{equation}
This prevents coils from interlinking.
    \item $J_\kappa$, the L2 curvature objective:
    \begin{equation}
        J_\kappa = \frac{1}{2}\sum_{i = 1}^{N_\text{coil}} \int_{C_i} \text{max}(\kappa - \kappa_0, 0)^2 ~dl.
    \end{equation}
Here, the target curvature $\kappa_0$ is set to the maximum curvature measured in the W7-X coil set.
\end{enumerate}

\printbibliography

\end{document}